\documentclass[prd,twocolumn,showpacs]{revtex4}
\usepackage{url}
\usepackage[bookmarks=false]{hyperref}
\usepackage{epsfig}
\usepackage{amsmath}
\usepackage{amssymb}
\usepackage[pdftex]{color}

\usepackage{listings}
\begin{document}
  
  \title{CUTE solutions for two-point correlation functions from large
         cosmological datasets}
  \author{David Alonso$^{1}$\thanks{E-mail: david.alonso@uam.es}}
  \affiliation{ $^{1}$Instituto de F\'isica Te\'orica UAM-CSIC,
                Universidad Aut\'onoma de Madrid, 28049 Cantoblanco,
                Spain}  \date{\today}

  \begin{abstract}
    In the advent of new large galaxy surveys, which will produce
    enormous datasets with hundreds of millions of objects, new 
    computational techniques are necessary in order to extract from
    them any two-point statistic, the computational time of which grows
    with the square of the number of objects to be correlated.
    Fortunately technology now provides multiple means to massively
    parallelize this problem. Here we present a free-source code
    specifically designed for this kind of calculations. Two
    implementations are provided: one for execution on shared-memory
    machines using OpenMP and one that runs on graphical processing
    units (GPUs) using CUDA. The code is available at
    \url{http://members.ift.uam-csic.es/dmonge/CUTE.html}.
  \end{abstract}
  \pacs{98.80.-k\hspace{\stretch{1}}IFT-UAM/CSIC-12-31}
  \date{\today}
  \maketitle  

  \section{Introduction}
  The last decades have seen an increasing interest in galaxy surveys
  as a means of studying the late-time evolution of the Universe.
  Forthcoming galaxy surveys, such as DES \citep{DES}, BigBOSS
  \citep{sch} or Euclid \citep{lau}, will map large regions of the
  sky ($\mathcal{O}(10^{3-4})$ sq-deg) to redshifts $z>1$ yielding
  catalogs containing hundreds of millions of objects.

  The spatial distribution of these objects on different scales
  contains invaluable information that could help clarify many open
  problems in cosmology and astrophysics, such as the nature of dark
  matter and dark energy or the presence of primordial
  non-Gaussianities in the density field. One of the simplest
  observables that can be estimated to quantify the clustering of
  matter on different scales is the two-point correlation function
  (2PCF hereon, see section \ref{sec:2pcf}). Its estimation is based
  on counting pairs of objects separated by a given distance measure,
  and therefore its computational time grows with the square of the
  number of objects in the catalog. Hence, when
  $\mathcal{O}(10^{14-16})$ pairs must be considered, a simplistic
  serial approach is too slow for the full-scale problem, and, besides
  using some simplifying approximation, the only viable solution
  becomes parallelising the calculation. In this sense modern
  graphical processing units (GPUs) provide the means to perform many
  operations in parallel on a large number (hundreds) of cores with a
  moderate clock frequency for a comparatively cheap price. Another
  approach is using a relatively smaller number of high-frequency CPU
  cores both in shared or distributed memory machines.

  Here we present a {\tt CUTE} (Correlation Utilities and Two-point
  Estimation), a free open-source code that estimates different kinds
  of two-point correlations from discrete cosmological catalogs
  using various speed-up techniques. 

  \section{The two-point correlation function(s)} \label{sec:2pcf}
  The three-dimensional 2PCF $\xi({\bf r})$ of a set of discrete
  points in $\mathbb{R}^3$ represents the excess probability of
  finding two of them inside two small volumes $dV_1$ and $dV_2$
  separated by ${\bf r}$ \citep{pe}:
  \begin{equation}
    \langle dP\rangle=\bar{n}[1+\xi({\bf r})]\,dV_1\,dV_2.
  \end{equation}
  When this point distribution comes from a Poisson process based on
  an underlying random density field $\delta({\bf x})$, the field's
  2PCF $\xi_{\delta}({\bf r})\equiv\langle\delta({\bf x})
  \delta({\bf x}+{\bf r})\rangle$ is directly related to that of the
  point distribution \citep{mar_saa}. Note that even though, in
  principle, the two-point correlation should depend on the
  positions of both points, ${\bf r}_1$ and ${\bf r}_2$, for
  homogeneous fields the only dependence is on the separation
  between them ${\bf r}\equiv{\bf r}_1-{\bf r}_2$.

  \subsection{Types of correlation functions}
  \begin{figure}
    \centering
    \includegraphics[height=0.35\textheight]{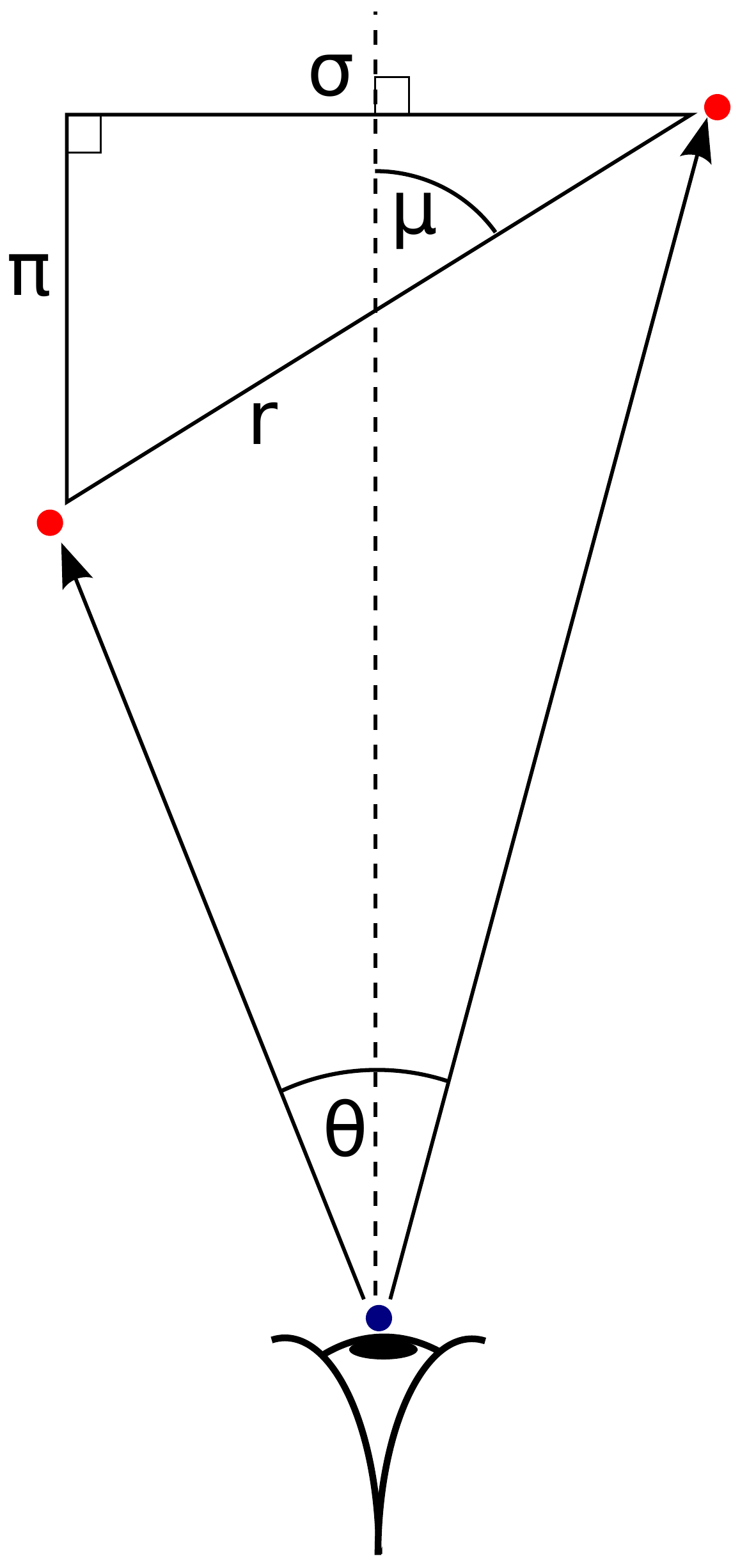}
    \caption{Definition of the different coordinate conventions used.}
    \label{fig:coords}
  \end{figure}

  \begin{itemize}
    \item {\bf The 3-D correlation function $\xi(r,\mu)$ and
      $\xi(\sigma,\pi)$.} Different observational effects, such as
      redshift-space distortions or errors in the observed redshifts,
      transform what would otherwise be an isotropic 2PCF into a
      function that behaves differently along the line of sight and
      in the transverse direction. Two coordinate systems are widely
      used in the literature: the $\sigma - \pi$ and $r - \mu$ schemes
      (see fig. \ref{fig:coords}), the relation between both being
      \begin{equation}
        \pi=r\,\mu,\,\,\,\,\,\sigma=\sqrt{r^2-\pi^2}.
      \end{equation}
      The $r - \mu$ scheme has the advantage that the usual multipole
      expansion is directly written in terms of these variables:
      \begin{equation}\label{eq:multip_exp}
        \xi(r,\mu)=\sum_l\xi_l(r)\,P_l(\mu),
      \end{equation}
      where $P_l$ are the Legendre polynomials. Note that at the linear
      level and in the plane-parallel approximation (i.e. the Kaiser
      formula \citep{kai}) only the first three even multipoles ($l=0,
      \,2,\,4$) contribute.
    \item {\bf The monopole $\xi_0(r)$.}
      The first element ($l=0$) in the expansion above is the
      angle-averaged correlation function or ``monopole'':
      \begin{equation}
        \xi_0(r)=\frac{1}{2}\int_{-1}^1d\mu\,\xi(r,\mu)
      \end{equation}
      This is the only non-zero contribution in the absence of redshift
      distortions.
    \item {\bf The radial correlation function
      $\xi_r(\bar{z},\Delta z)$.} Correlating only pairs of galaxies
      aligned with the line of sight, one computes the so-called radial
      correlation function, which can be made to depend locally only on
      the redshift difference $\Delta z$ between each pair of galaxies.
      This quantity is related to the three-dimensional 2PCF through
      \begin{equation}
            \xi_r(\bar{z},\Delta z)=\xi(\pi(\bar{z},\Delta z),\sigma=0),
          \end{equation}
          whith
          \begin{equation}
            \pi(\bar{z},\Delta z)\simeq \frac{c\,\Delta z}{H(\bar{z})},
          \end{equation}
          where $\chi(z)$ is the radial comoving distance to redshift $z$.
    \item {\bf The angular correlation function $w(\theta)$.}
          The angular correlation function is the 2PCF of the density contrast field projected on
          the sphere
          \begin{align}
            &w(\theta)\equiv\langle\delta_s(\hat{\bf n}_1)\delta_s(\hat{\bf n}_2)\rangle,
            \,\,\,\,\cos\theta\equiv\hat{\bf n}_1\cdot\hat{\bf n}_2,\\
            &\delta_s(\hat{\bf n})\equiv \int\,dz\, \phi(z)\,\delta(r(z)\,\hat{\bf n}),
          \end{align}
          where $\phi(z)$ is the redshift selection function. The angular correlation function is
          related to $\xi(r,\mu)$ by
          \begin{align}\label{eq:w_theta}
            & w(\theta)=\int\,dz_1\,\phi(z_1)\int\,dz_2\,\phi(z_2)\,\xi(r(z_1,z_2,\theta),
                        \mu(z_1,z_2,\theta)) \\ \nonumber
            & r(z_1,z_2,\theta)=\sqrt{\chi^2(z_1)+\chi^2(z_2)-2\,\chi(z_1)\,\chi(z_2)\,\cos\theta},
                        \\\nonumber
            & \mu(z_1,z_2,\theta)=\frac{|\chi^2(z_1)-\chi^2(z_2)|}{\sqrt{(\chi^2(z_1)+
                        \chi^2(z_2))^2-4\,\chi^2(z_1)\chi^2(z_2)\cos^2(\theta) }}
          \end{align}
  \end{itemize}

  \subsection{Estimating the 2PCF from discrete data}
  As we have said, the two-point correlation function can be understood as the excess probability
  of finding two objects separated by a given distance with respect to a random distribution, and
  therefore:
  \begin{equation}
    1+\xi = \frac{N_p^d(r)\,dr}{N_p^r(r)\,dr}
  \end{equation}
  where $N_p^d\,dr$ is the number of pairs separated by $r\pm dr/2$ in the data, and $N_p^r(r)\,dr$
  is the number of pairs that one would expect for a random distribution. The numerator can be
  easily calculated as
  \begin{equation}
    N_p^d(r)\,dr = \sum_{i=1}^N \sum_{j\neq i} \Theta(r-dr/2<|{\bf x}_i-{\bf x}_j|<r+dr/2),
  \end{equation}
  where $N$ is the total number of objects, $\Theta$ is 1 whenever its argument is true and 0
  otherwise, and we have explicitly avoided counting self-pairs. If the catalog had no boundaries,
  the number of random pairs could easily be estimated as
  \begin{equation}
    N_p^r(r) = \frac{N^2}{V}v(r),
  \end{equation}
  where $v(r)\simeq 4\pi\,r^2\,dr$ is the volume of a spherical shell of radius $r$ and thickness
  $dr$.
  
  As we have said, this is can only be done if the catalog has no boundaries. Effectively this
  true in the case of an N-body simulation, where a sphere that lies partly ouside the simulation
  box can be ``wrapped around'' due to the periodic boundary conditions. Thus, in this case a
  possible estimator is
  \begin{equation}\label{eq:estim_box}
    1+\xi(r)\equiv\frac{V}{N^2v(r)}\sum_{i,j\neq i} \Theta(r-dr/2<|{\bf x}_i-{\bf x}_j|<r+dr/2),
  \end{equation}
  
  However, when calculating the correlation function from a point distribution with complicated 
  boundaries, as is usually the case in a galaxy survey, several observational dificulties arise:
  e.g. different parts of the sky may have been mapped to different depths and the radial
  distribution of objects (selection function) is never uniform. The most usual technique to deal with
  these issues is to compare the data catalog with catalogs made of randomly distributed objects
  that also contain these artificial effects. In this case the 2PCF can be naively estimated as
  \begin{equation}
    \xi_N=\frac{N_r(N_r-1)}{N_d(N_d-1)}\frac{DD}{RR}-1
  \end{equation}
  where $N_d$ and $N_r$ are the number of points in the data and random catalogs respectively and
  $DD$ and $RR$ are histograms containing the counts of pairs of objects found separated by a given
  distance in each catalog. It has been shown \citep{lan_sza} that the variance of this estimator can
  be minimized, and its ability to cope with boundary conditions can be enhanced, by making use of
  the cross-correlation of random and data objects, $DR$. The most widely used estimator is the one
  proposed by Landy \& Szalay \citep{lan_sza}:
  \begin{equation}
    \xi_{LS}=\frac{\frac{N_r(N_r-1)}{N_d(N_d-1)}DD-\frac{N_r-1}{N_d}DR+RR}{RR}.
  \end{equation}
  See \citep{ker} for a thorough comparison of different estimators.

  The most delicate part of the estimation is in fact being able to generate the random catalogs
  correctly: the background spatial distribution, both in angles and redshift (i.e. the one-point
  function), of random objects must be exactly the same as in the data. Hence, all observational
  effects that affect the spatial distribution must be correctly reproduced by the random catalogs.
  Also, in order to minimize Poisson errors in $DR$ and $RR$, random catalogs should be generated
  with more particles than the data.

  \section{CUTE}
  {\tt CUTE} (Correlation Utilities and Two-point Estimation) is a free and open-source code for
  cosmological 2PCF estimation. {\tt CUTE} is written in C and, in the current public version, comes with
  two implementations: one parallelized for shared-memory machines using OpenMP and one ({\tt CU\_CUTE})
  that performs the correlations in a GPU using NVidia's CUDA architecture. {\tt CUTE} calculates 4
  different correlation functions (3-D, monopole, angular and radial) with different binning schemes
  and speed-up techniques. Here we will explain the parallelization strategies followed by {\tt CUTE} and
  some details specific to each type of 2PCF. We refer the reader to the README file accompanying
  the latest public version of {\tt CUTE} for the operational options and compilation instructions of the
  code. In this section we assume some basic knowledge of parallel computing with OpenMP and CUDA by the
  reader.

  It must be noted that there exist two other codes \citep{pon,bar}, recently made public,
  designed to compute angular correlation functions with GPUs. As in the case of {\tt CUTE} the speed-up
  factor (about $10^2$) gained by these codes through the use of graphical devices for
  parallelization clearly makes it worth the effort of adapting CPU algorithms to run on GPUs.

  \subsection{Serial approach}
  Once the random catalog has been produced, $DD$, $DR$ and $RR$ are computed by autocorrelating or
  crosscorrelating each pair of catalogs. In a serial code this algorithm is extremely simple,
  involving one loop over each catalog and performing 3 operations in each iteration: calculating
  the distance between each pair of objects, determining the bin corresponding to that distance and
  increasing the histogram count on that bin. The corresponding C-code would be:
  \begin{lstlisting}
int histogram[nbins];
for(i=0;i<np1;i++) {
  for(j=0;j<np2;j++) {
    //Calculate distance between two objects
    double dist=get_dist(x1[i],y1[i],z1[i],
                         x2[j],y2[j],z2[j]);
    //Calculate bin number
    int ibin=bin_dist(dist);
    //Increase histogram count
    histogram[ibin]++;
  }
}
  \end{lstlisting}
  As we said before, these two nested loops make this an $N^2$ problem (to be precise, an
  {\tt n1*n2} problem), whose computational time will grow very fast as we increase the size of the
  catalogs. At this point parallelization or/and some kind of fast approximate method are desirable,
  if not compulsory. In section \ref{ssec:paral} we will describe the parallelization strategies used
  by CUTE, which complicate this simple algorithm. Other speed-up techniques used by the code are
  explained in section \ref{ssec:neigh}, and some especific details of each type of 2PCF are given in
  section \ref{ssec:2pcfs}.

  \subsection{Parallelization with CUDA and OpenMP}\label{ssec:paral}
  \subsubsection{Multicore shared-memory machines and OpenMP}
  OpenMP \footnote{\url{http://www.openmp.org}} is an API that gives support for parallel programming in
  shared-memory platforms. Once a parallel execution block is opened, the programmer can define private
  (one independent copy per core) or shared (common) variables and easily divide {\tt for} loops
  between all available cores. For a thorough review of the different features of OpenMP see
  \citep{cha}. The serial code above takes the following form when parallelized with OpenMP:
  \begin{lstlisting}
int histogram[nbins];
int histo_thread[nbins];
#pragma omp parallel default(none) \
  private(hthread) shared(...) {
  //Initialize private histograms
  for(i=0;i<nbins;i++)
    histo_thread[nbins]=0;
#pragma omp for    //Parallelize loop
  for(i=0;i<np1;i++) {
    for(j=0;j<np2;j++) {
      //Calculate distance between two objects
      double dist=get_dist(x1[i],y1[i],z1[i],
                           x2[j],y2[j],z2[j]);
      //Calculate bin number
      int ibin=bin_dist(dist);
      //Increase histogram count
      histo_thread[ibin]++;
    }
  }
#pragma omp critical {
  //Add private histograms
  for(i=0;i<nbins;i++)
    histogram[i]+=histo_thread[i]
}
}
  \end{lstlisting}
  The strategy in this case to is to declare one private histogram per execution thread that will
  store that thread's pair counts. The first loop is then divided between all available threads and
  finally all partial histograms are added, avoiding read/write collisions, into the final shared
  one. As can be seen, parallelization with OpenMP is effortless, only requiring a few extra lines
  of code.

  \subsubsection{Graphics cards and CUDA}\label{sec:CUDA}
  A GPU (Graphical Processing Unit) is a specialized piece of hardware designed for fast massively
  parallel manipulation of memory addresses. As their name suggests, GPUs are mainly intended for
  image building and processing, however their highly parallel structure makes them ideal for
  intensive numerical computation, providing a relatively cheap flop/s (floating-point operations
  per second). Hence in the last years GPUs have found their way into different branches of
  scientific research, computational cosmology being one of them \citep{bed,nak}
  Initially the main difficulty when trying to use GPUs for scientific computing was
  the programming of the numerical algorithm, since using the standard APIs meant that data had to
  be disguised as pixel colors and some mathematical operations had to be encoded as graphics
  rendering. However lately a few programming models for GPUs have seen the light of day
  \footnote{\url{http://www.nvidia.com/object/cuda\_home.html}} \footnote{\url{http://www.khronos.org/opencl/}}
  that make general purpose computing on GPUs (GPGPU) a lot easier. Of these we have
  chosen Nvidia's CUDA \citep{nic} for its syntactic simplicity.

  Two main complications arise when one tries to adapt a code to execute on a GPU. First, in a
  massively parallel environment one must take great care to avoid race conditions due to
  simultaneous memory read/write processes by different threads. Second, unlike in a multi-CPU
  machine, the amount of memory ``per thread'' available in a GPU is very limited, presently of the
  order of a few GB for hundreds of processors. Besides these, there are other more subtle concerns,
  such as intra-warp communication or the presence of different types of cached and uncached
  memory in the GPU, the correct use of which may enhance dramatically the code's performance. In
  summary, correctly parallelising a code with CUDA is not as straightforward as it is with OpenMP,
  and in some cases it may not be worth the effort. For an introduction to CUDA and its many
  features see \citep{san_kan}.

  Implementing the serial algorithm above in CUDA would involve executing the following
  {\tt \_\_device\_\_} function in every thread in parallel:
  \begin{lstlisting}
__shared__ int histo_thread[nbins];
int stride=blockDim.x*gridDim.x;
//Initialize shared histogram
histo_thread[threadId.x]=0;
__syncthreads();
//Correlate
for(i=0;i<np1;i++) {
  int j=threadIdx.x+blockIdx.x*blockDim.x;
  while(j<np2) {
    //Calculate distance between two objects
    double dist=get_dist(x1[i],y1[i],z1[i],
                         x2[j],y2[j],z2[j]);
    //Calculate bin number
    int ibin=bin_dist(dist);
    //Increase histogram count
    atomicAdd(&(histo_thread[ibin]),1);
    //Increase second index by stride
    j+=stride;
  }
}
//Add block histograms
__syncthreads();
atomicAdd(&(histogram[threadIdx.x]),
          histo_thread[threadIdx.x]);
  \end{lstlisting}
  As before, we have divided one of the loops (this time the second one) among all the execution
  threads. The first difference with respect to OpenMP that we can see inmediately is that, due to
  the limited amount of memory of the GPU, we can only declare one partial histogram per block, and
  not per thread. To do this we declare it as a variable in shared memory, which also has the
  advantage of having a lower latency than global memory. This introduces a new complication, since
  now all threads in a block will try to add their pair counts to the same histogram. This has to
  be done avoiding race conditions by using the CUDA {\tt atomicAdd()} function. This is in fact
  the bottleneck of any algorithm involving histograms in CUDA (especially if the distribution
  under study is very degenerate), since many threads may have to remain idle while waiting for
  other threads to update their histogram entries. The best way to palliate this problem is to use
  at most as many threads per block as histogram bins (in fact, note that the algorithm above will
  only work when using as many threads as histogram bins, however it can be easily extended to more
  general cases). Finally all the partial block histograms are summed up into the global histogram
  in an ordered manner using again {\tt atomicAdd()}. The fact that {\tt CUTE} uses CUDA atomic functions
  such as {\tt atomicAdd()} means that it will only run on GPUs that support atomic operations
  (namely compute capability 2.0 or higher). There exist general algorithms for histograms that work
  on any CUDA-enabled device \citep{pod,sha_ken}, however no performance improvement was observed
  with respect to using {\tt atomicAdd()}. Furthermore, the method above reduces the use of shared
  memory for histograms to a minimum, allowing its use for other purposes.

  \subsection{Neighbor searching}\label{ssec:neigh}
  \begin{figure}
    \centering
    \includegraphics[width=0.45\textwidth]{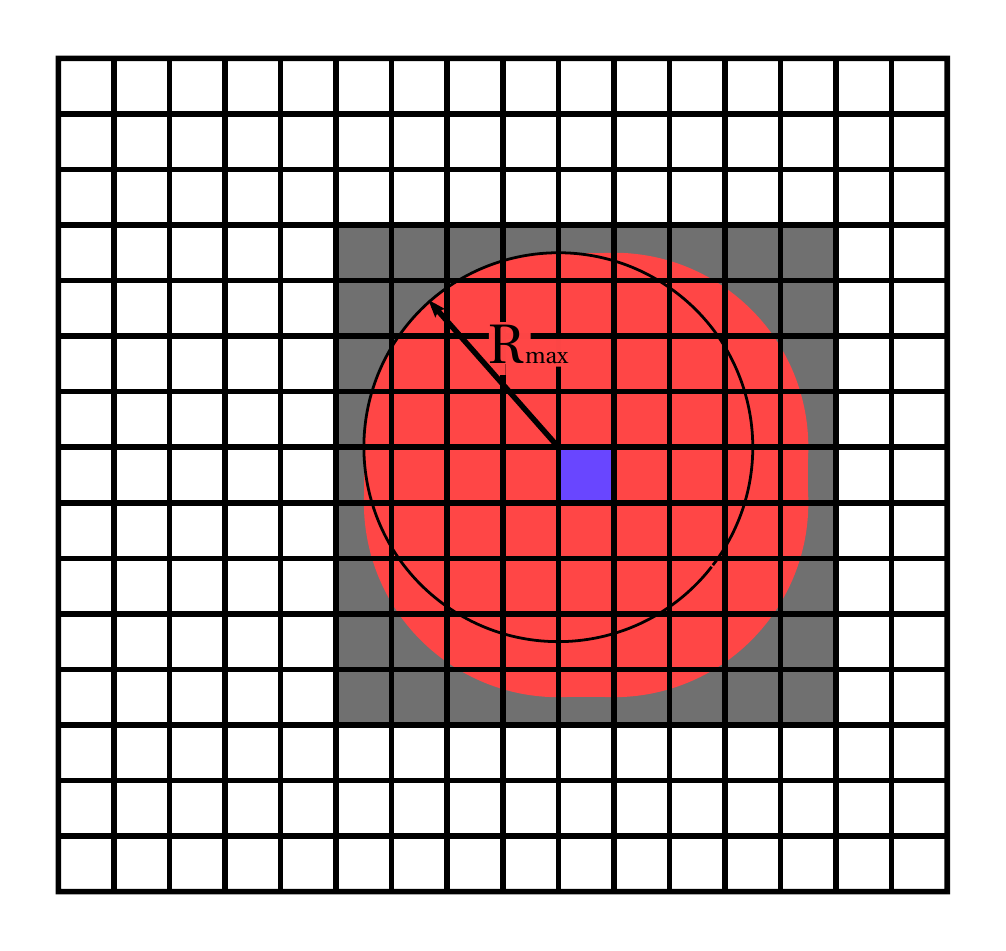}
    \includegraphics[width=0.45\textwidth]{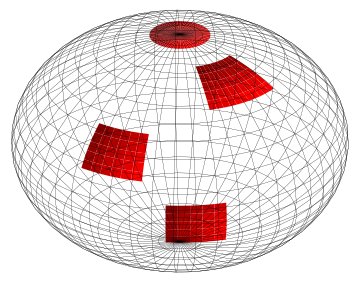}
    \caption{Illustration of the main neighbor-searching technique used by {\tt CUTE}. In the
             three-dimensional case (top panel), the catalog is covered by cubical cells. Around
             each cell $C_i$ (blue), a larger cube is drawn (gray), that safely contains all
             spheres of radius $R_{\rm max}$ centered within $C_i$ (red). Neighbors of the objects
             within $C_i$ are only searched for in the gray region. The bottom panel shows the similar
             neighbor-searching regions used on the sphere for the calculation of the angular
             2PCF. In this case the shape of the region is different depending on the position of
             the central pixel.}
    \label{fig:neighbors}
  \end{figure}
  Often the maximum scale to which we want to calculate the 2PCF is significantly smaller than the
  size of our data. In this case, calculating the relative distance between particles that are
  further away than this maximum scale is useless, and therefore should be avoided. However, how
  can we determine which pairs to avoid without actually calculating their distances? {\tt CUTE} makes
  use of different approaches to minimize the amount of useless pair counts in an efficient way.
  The main strategy described here is very similar in the three-dimensional case (for the 3-D
  and monopole 2PCFs) and on the sphere (for the angular correlation), however they differ slightly
  in the details.
  
  In the three-dimensional case, a box encompassing the whole catalog is first determined and
  divided into cubical cells. To each cell we associate the positions of all the objects that fall
  inside it. Assuming that the maximum distance we are interested in is $R_{\rm max}$ and that
  the cell size is $a$, we draw a cube of $2\,\lfloor R_{\rm max}/a\rfloor+1$ cells per side around
  each cell $C_i$ (here $\lfloor b\rfloor$ denotes the integer part of $b$). This guarantees that
  we can draw spheres of radius $R_{\rm max}$ around any point in $C_i$ and that these spheres will
  all lie inside the cube (see the top panel in figure \ref{fig:neighbors}) . Thus we can correlate
  all the objects inside $C_i$ with the objects in all the other cells inside the cube and safely
  ignore all other objects. The efficiency of this method depends largely on the number density of
  the catalog, the range of scales of interest, and the number of cubical cells used.

  In the spherical case a similar approach is used. Let us define a spherical cube as a region
  of the sphere with constant limits in spherical coordinates, i.e. a region with 
  $\phi_0<\phi<\phi_f$ and $\cos\theta_f<\cos\theta<\cos\theta_0$. It is easy to prove
  that the spherical cube containing a spherical cap of radius $\theta_{\rm max}$
  centered at the point $(\theta,\phi)$ in spherical coordinates has sides of length:
  \begin{align}\nonumber
    &\Delta(\cos\theta) = \cos(\theta-\theta_{\rm max})-\cos(\theta+\theta_{\rm max}),\\
    &\Delta(\phi) = \frac{\sqrt{\cos^2\theta_{\rm max}-\cos^2\theta}}{\sin\theta}.
  \end{align}
  Now, in the spherical case we can use pixels defined as small spherical cubes instead of the
  cubical cells of the three-dimensional case. Then the result we have just quoted can be used
  to define a spherical cube of pixels centered at a given one that safely contains all particles
  within an angular distance $\theta_{\rm max}$ of any particle in the central pixel (see the
  bottom panel in figure \ref{fig:neighbors}). Once this is done, the same procedure is followed
  as in the three-dimensional case.
  
  Another more sophisticated and very popular technique to discard unnecessary correlations is
  the so-called $k$-Tree method. For a thorough description of this method, see \cite{moore}.

  \subsection{Specifics of the 2PCFs}\label{ssec:2pcfs}

  \begin{figure}
    \centering
    \includegraphics[width=0.45\textwidth]{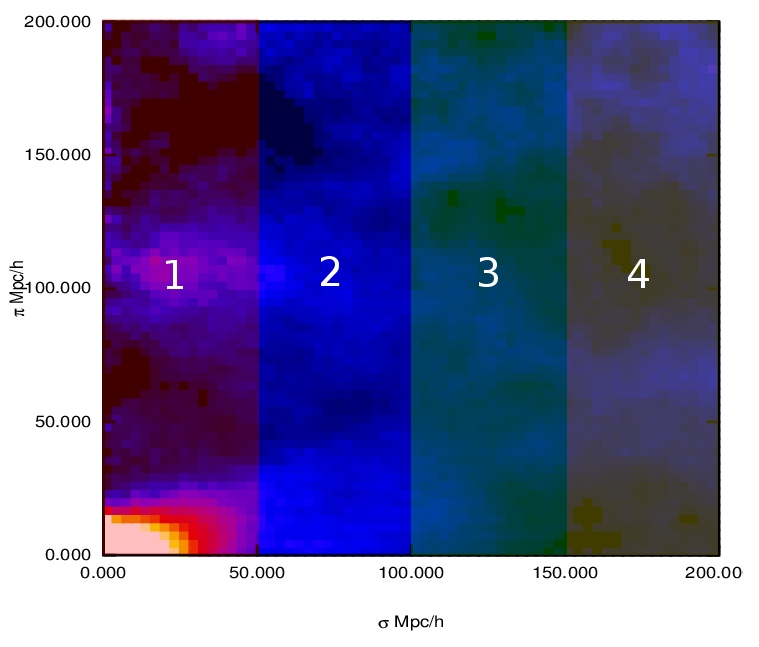}
    \caption{Illustration of the method used to calculate the 3-D correlation function in CUDA.
             Since the whole 2-dimensional $128\times128$ bin histogram cannot be fit inside the
             device's shared memory, it is split into smaller ones, which are filled separately.
             Although the catalogs have to be correlated more than once, the bottleneck caused by
             atomic operations is largely mitigated by the higher number of histogram bins.}
    \label{fig:3d_method}
  \end{figure}

  In the previous section we have described the general strategy followed to parallelize the
  calculation of any 2PCF with OpenMP and CUDA. However each of the 2PCFs detailed in
  section \ref{sec:2pcf} requires a different treatment of the data and maybe allows for different,
  more optimal, approaches. We give the details specific to each of these types here.

  \subsubsection{Radial correlation function}
  As was said in section \ref{sec:2pcf} the radial 2PCF is calculated by correlating pairs of
  aligned objects and binning them according to their relative redshift difference $\Delta z$.
  Spherical cubes are used by {\tt CUTE} to quickly find pairs of galaxies subtending an angle
  smaller than some maximum aperture, which defines aligned pairs. For reasonable apertures
  ($\lesssim1^o$) the number of pairs to correlate is relatively small. Hence, since the
  computational time in this case is not an issue, there is no need for massive parallelization,
  and radial correlation functions are only supported by {\tt CUTE} in its OpenMP version.

  \subsubsection{Angular correlation function}\label{sec:ang_corr}
  For the calculation of angular 2PCFs {\tt CUTE} projects all objects in the catalog into the unit
  sphere and correlates pairs of objects according to their angular separation $\theta$ (see
  figure \ref{fig:coords}), which is used as a distance measure. Two complementary speed-up
  techniques can be used by {\tt CUTE} in this case: if one is not interested in extremely small
  angular scales one can create a pixel map from the catalog and then correlate the pixels
  (weighting each of them by the number of objects that fall inside it). This may effectively
  reduce the number of objects that must be correlated by an order of magnitude and therefore
  reduce the computational time by a factor of 100. Also, in the calculation of the angular
  separation, the arc-cosine of the scalar product of two position vectors must be estimated.
  Calculating the arc-cosine is a very time-consuming operation, and, if one is not interested in
  very large angular scales, the following approximation can be used,
  \begin{equation}\nonumber
    \arccos(1-x)\sim\sqrt{2\,x+\frac{1}{3}x^2+\frac{4}{45}x^3}
  \end{equation}
  which is precise to 1 part in $10^{-4}$ for angles below 40$^o$ and reduces the computational
  time by a factor $\sim 2$. Both these time-saving techniques can be swiched on or off in {\tt CUTE}
  by the user.

  \subsubsection{3-D correlation function}
  The main difference in the calculation of the 3-D correlation function with respect to the other
  2PCFs is that pairs are binned in 2-dimensional histograms, according to their $(r,\mu)$ or
  $(\pi,\sigma)$ separations. This does not introduce any relevant changes in the OpenMP
  implementation, as long as the the amount of shared memory is large enough to accomodate one
  private 2-D histogram per thread, however it does matter when adapting the code to CUDA. The
  reason is that currently the amount of shared memory per block in GPUs is limited to 48 kB, which
  is too little to allocate, for example, a $128\times128$ array of long integers. The solution to
  this problem chosen for {\tt CUTE} is explained in fig. \ref{fig:3d_method}: the catalogs are
  correlated several times, each time binning only pairs whose separation in one of the two
  coordinates is within a given range, until the whole histogram is filled. Thus we can declare
  smaller 2-D histograms in shared memory, and even though the catalogs must be correlated several
  times, the histogram-filling bottleneck mentioned in section \ref{sec:CUDA} is largely alleviated
  by the higher number of histogram entries, thus conserving a reasonable computational time (see
  section \ref{sec:speed}).

  \subsubsection{The monopole in a box}
  In its current public version, {\tt CUTE} has a companion program, {\tt CUTE\_box}
  that calculates the correlation function from data inside a cubical box with periodic
  boundary conditions. In this case only the calculation of the isotropic 2PCF (the monopole)
  is supported, using two different types of algorithms:
  \begin{itemize}
   \item {\bf Particle-based algorithms}. In this case {\tt CUTE} calculates the 2PCF
     from the pair counts using the estimator in equation \ref{eq:estim_box}. As we have
     discussed, no random catalog is needed because of the periodic boundary conditions. Two
     different types of neighbor-searching algorithms are supported: cubical cells and 
     $k$-Trees.
   \item {\bf Density grid}. This algorithm is similar to the use of pixels to accelerate
     the calculation of the angular correlation function. In this case the particle content
     of the catalog is first interpolated to a grid and the overdensity field $\delta$ is
     estimated at every grid point using a TSC algorithm. Then pairs of grid points are
     correlated, and a weight $\delta_i\,\delta_j$ is given to each pair. The correlation 
     function is estimated as
     \begin{equation}
       \hat{\xi}(r)=\langle \delta({\bf x})\delta({\bf x}+{\bf r})\rangle,
     \end{equation}
     where the average is taken over all pairs of grid points separated by a distance $r$.
     Due to the simplicity of a regular grid, it is trivial to search for the neighboring
     grid points, and since the relative distances between neighboring points are the same
     everywhere, these relative distances only have to be calculated once. As a result of
     this, this method is usually the fastest, however it will only yield reliable results
     down to the scale of the grid.
  \end{itemize}

  \section{Performance}\label{sec:speed}

  \begin{table*}
    \begin{center}
      \begin{tabular}{|c|c|c|c|}
        \hline
        & Name & Description & \#cores \\
        \hline
             & Sequential & Intel Core i7-2620M             & 1 core ($\equiv1$ thread)      \\
        CPUs & Laptop-MP  & Intel Core i7-2620M             & 2 cores ($\equiv4$ threads)    \\
             & Server-MP  & Intel MP NEHALEM-EX ($\times8$) & 80 cores ($\equiv160$ threads) \\
        \hline
        GPUs & Laptop-GPU & NVIDIA NVS 4200M                & 48 CUDA cores                  \\
             & Server-GPU & NVIDIA TESLA C2070 FERMI        & 448 CUDA cores                 \\
        \hline
      \end{tabular}
    \end{center}
    \caption{Different devices in which {\tt CUTE} has been tested: a single CPU core, a dual core,
             a multi-core shared-memory machine (160 threads), an ordinary graphics card and 
             a high-end GPU.}\label{tab:devices}
  \end{table*}

  \begin{table*}
    \begin{center}
      \begin{tabular}{|c|c|c|c|c|c|}
        \hline
        Platform   & $T(\xi(r))$ & $T(\xi_{\rm log}(r))$ & $T(w(\theta))$ & $T(w_{\rm pix}(\theta))$ & $\xi(\sigma,\pi)$ \\
        \hline
        Sequential &    877      &        5230           &     1374       &       21                 &     2238          \\
        Laptop-MP  &    389      &        2676           &     628        &       5.3                &     1064          \\
        Laptop-GPU &    113      &        185            &     283        &       6.2                &     297           \\
        Server-MP  &     25      &        52             &     32         &       0.51               &     50            \\
        Server-GPU &     13      &        20             &     22         &       0.46               &     27            \\
        \hline
      \end{tabular}
    \end{center}
    \caption{Computational times ellapsed, for each of the 5 platforms listed in table \ref{tab:devices},
             during the calculation of 5 different 2PCFs: monopole ($\xi(r)$), monopole with logarithmic
             binning ($\xi_{\rm log}(r)$), angular ($w(\theta)$), angular with pixels of resolution
             $\Delta\Omega\equiv5\times10^{-3}$ sq-deg ($w_{\rm pix}(\theta)$) and 3-D ($\xi(\sigma,\pi)$).
             Times are in seconds and correspond to the calculation of the $DR$ histogram (the full 
             calculation of the 2PCF should take 2-3 times longer).}\label{tab:times}
  \end{table*}

  \begin{figure}
    \centering
    \includegraphics[width=0.45\textwidth]{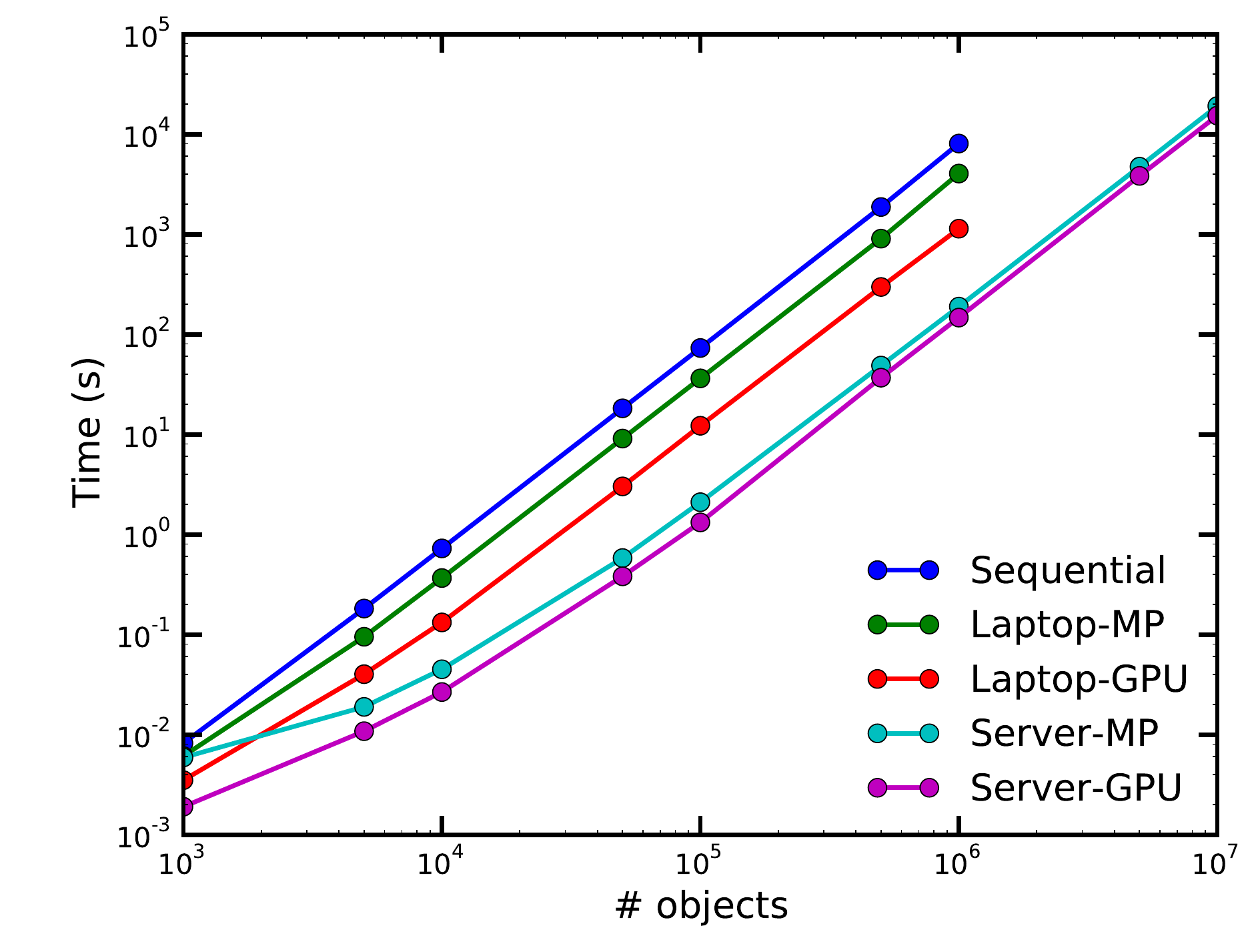}
    \caption{Computational times employed by different devices to compute the monopole 2PCF of
             catalogs of different sizes. A speed-up factor of $\mathcal{O}(100)$ can be gained by
             using a high-end GPU with respect to a sequential approach on a high-end CPU. Even
             with a regular gaming GPU the increase in speed is substantial ($\mathcal{O}(10)$).
             The different devices are described in table \ref{tab:devices}.}
    \label{fig:timings}
  \end{figure}

  We have tested {\tt CUTE}'s performance in terms of computational time by running it on platforms 
  with different capabilities, listed in table \ref{tab:devices}. The serial version was tested by
  running {\tt CUTE} on a single CPU core. We also tested the OpenMP version on a dual-core laptop
  and on a large shared-memory machine with 80 cores. The CUDA version has been tried on a regular
  graphics card in a laptop and on a high-end GPU. All the computational times quoted in this
  section correspond to tests performed without any of the neighbor-searching techniques described
  in section \ref{ssec:neigh} in order to provide a clearer comparison between platforms, and
  they should therefore be understood as the worst-case scenario. As we have noted, the use of
  these strategies may improve the computational time significantly with respect to a more naive
  approach (even by orders of magnitude). However, this improvement depends largely on the number
  density of the data and the scales of interest.
  
  For this test the monopole 2PCF was calculated for catalogs of different sizes in the range
  $10^3-10^7$. The computational times for one single correlation (i.e. just calculating, for
  example, $DR$) in the 5 different platforms are plotted in figure \ref{fig:timings}. As
  expected, using GPUs or parallelising the computation on several CPU cores improves the
  code's speed by a factor 10 - 100, even using a regular video-game graphics card. The
  ellapsed times were measured using OpenMP and CUDA timing functions, since these give
  the most accurate estimate of the time spent doing the actual correlation.

  For completeness we have also listed in table \ref{tab:times}
  the computational times taken by the 5 different devices to calculate different correlation 
  functions. The dataset used for this exercise is a subset of one of the mock catalogs
  provided by the MICE project \footnote{\url{http://maia.ice.cat/mice/}} \citep{fos}, 
  with $0^o<{\rm dec}<18^o$, $0^o<{\rm R.A.}<18^o$, $0.5<z<0.6$, containing $\sim 3\times10^5$
  particles. The 5 different correlation functions are:
  \begin{itemize}
    \item Monopole correlation function: linear binning for $r<100\,{\rm Mpc}/h$ and 256 bins.
    \item Monopole correlation function: logarithmic binning for $r<100\,{\rm Mpc}/h$ using 256 bins
          and 50 bins per decade.
    \item Angular correlation function: linear binning for $\theta<30^o$ and 256 bins. Calculated
          by brute-force.
    \item Angular correlation function: linear binning for $\theta<30^o$ and 256 bins. Calculated
          using pixels with resolution $\Delta\Omega\equiv5\times10^{-3}$ sq-deg.
    \item 3-D correlation function: binning in $(\pi,\sigma)$ on a $64\times64$-bin histogram.
  \end{itemize}

\begin{figure*}
    \centering
    \includegraphics[width=0.65\textwidth]{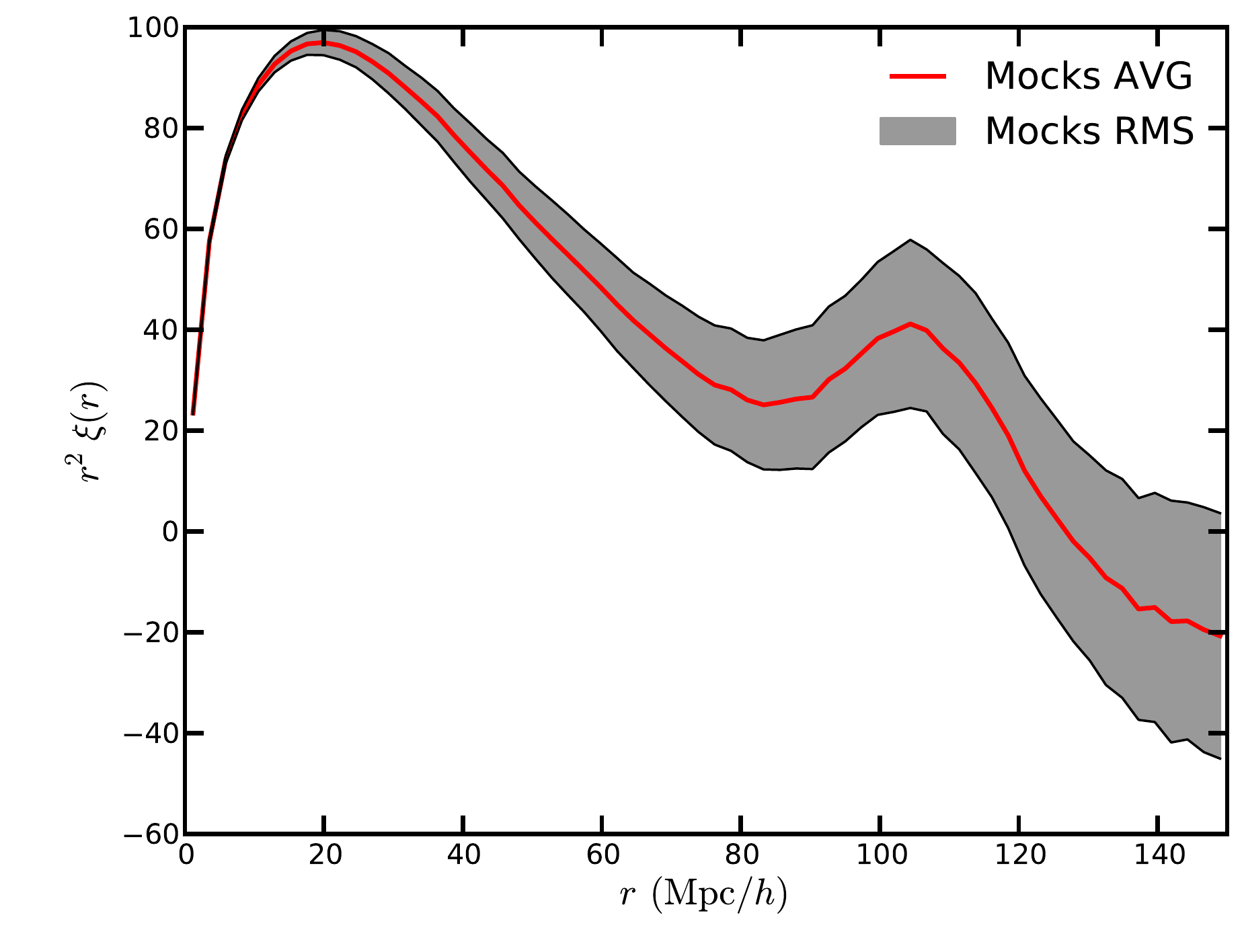}
    \caption{Monopole correlation function calculated from the PTHalo mock catalogs used in the analysis of the SDSS Ninth
             Data Release \cite{manera}. The solid red line shows the average correlation function, and the shaded area shows the
             1-$\sigma$ region around it, calculated as the r.m.s. over the 600 mocks. The monopole was calculated for
             the 600 mock catalogs in about 2 hours by running CUTE on the platform Server-MP (see table
             \ref{tab:devices}).}
    \label{fig:corrs1}
  \end{figure*}

\begin{figure*}
    \centering
    \includegraphics[width=0.65\textwidth]{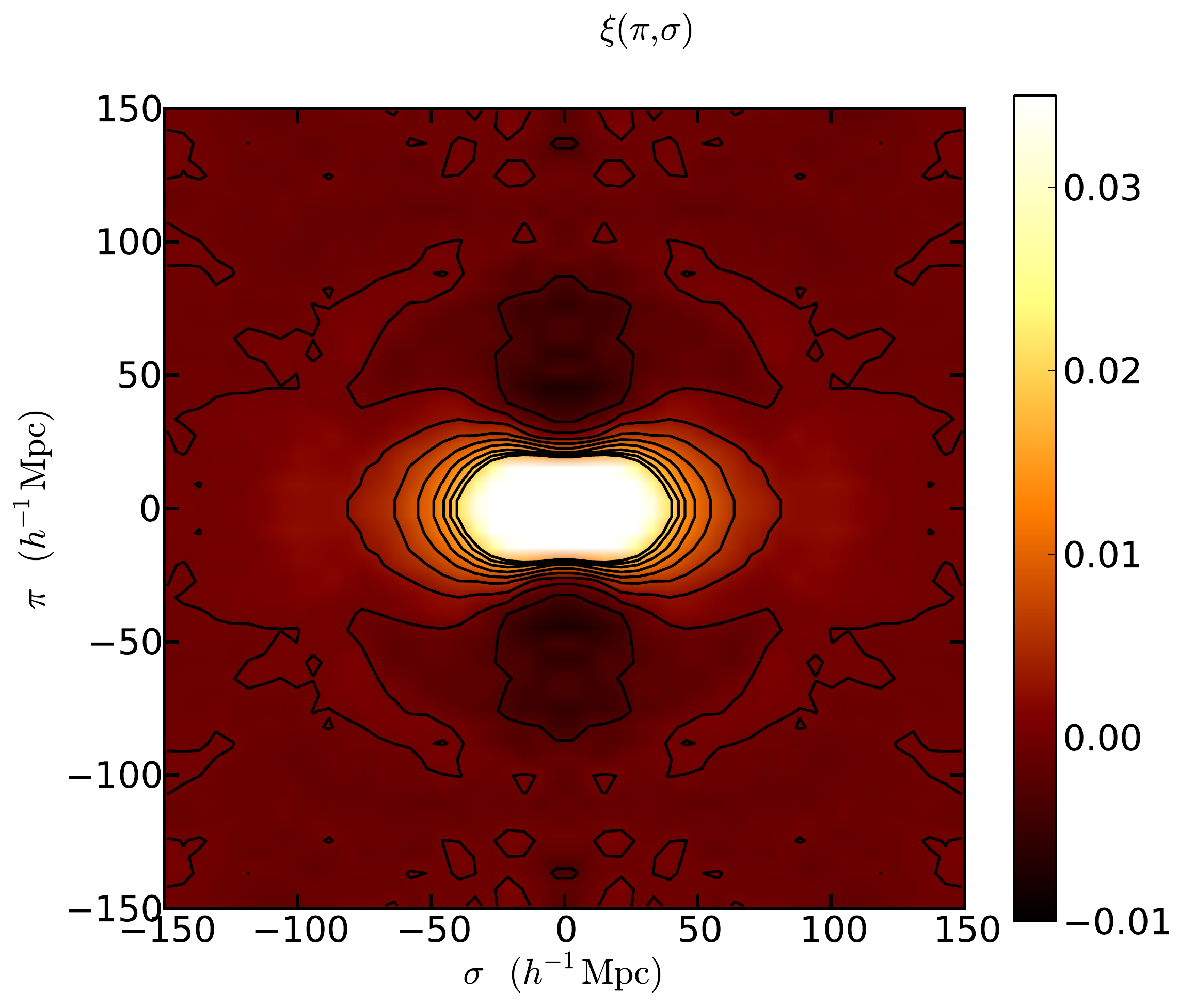}
    \caption{3-D correlation function calculated from a log-normal mock catalog. The catalog contained $\sim4.3\times10^7$
             objects in one octant of the sky between redshifts 0.45 and 0.75. The brute-force calculation took $\sim 10$
             hours on the platform Server-GPU (see table \ref{tab:devices}). Redshift-space distortions produce a squashing
             of the correlation function along the transverse direction and create a region of negative correlation along
             the line of sight. This was first noted in \cite{gaztacabre}.}
    \label{fig:corrs2}
  \end{figure*}

  Figures \ref{fig:corrs1} and \ref{fig:corrs2} show the output produced by CUTE for different kinds
  of correlation functions.

  \section{Summary}
  We have presented {\tt CUTE}, a parallel code for computing two-point correlation functions from
  cosmological catalogs. {\tt CUTE} has been optimized to run on shared-memory machines as well as
  graphical processing units. It can estimate the 3-D, monopole, radial and angular correlation
  functions from a set of data using different speed-up techniques and binning schemes. We
  have shown that great benefits in terms of computational speed can be gained by parallelising the
  algorithm on GPUs.

  The code is publicly available through our website \footnote{\url{http://members.ift.uam-csic.es/dmonge/CUTE.html}}.
  {\tt CUTE} is released under the GNU Public License (GPL).

  \section*{Acknowledgments}
  The author would like to thank Ignacio Sevilla, Miguel C\'ardenas and Rafael Ponce for their
  invaluable input and Alexander Knebe for useful suggestions and beta-testing. CUTE was initially tested
  on mock data kindly provided by the MICE collaboration. DA acknowledges
  support from a JAE-Predoc contract.

\end{document}